\documentclass{article}
\usepackage{spconf,amsmath,graphicx}
\usepackage{amssymb,amsmath,amsthm}
\usepackage{xifthen}
\usepackage{graphicx,xcolor}
\graphicspath{{figures/}}

\newcommand{\ing}[1]{\mathsf{#1}}
\newcommand{\Rn}[1]{\ifthenelse{\equal{#1}{}}{\mathbb{R}}{\mathbb{R}^{\ing{#1}}}}
\newcommand{\Cn}[1]{\mathbb{C}^{\ing{#1}}}
\newcommand{\Rset}[2]{\ifthenelse{\equal{#2}{1}}{\in \Rn{\ing{#1}}}{\in \Rn{\ing{#1} \times \ing{#2}}}}
\newcommand{\Cset}[2]{\ifthenelse{\equal{#2}{1}}{\in \Cn{\ing{#1}}}{\in \Cn{\ing{#1} \times \ing{#2}}}}

\newcommand{\var}[1]{\mathrm{#1}}

\newcommand{\ind}[2]{\ifthenelse{\equal{#2}{}}{\chi_{#1}}{\chi_{#1}\left( #2 \right)}}

\newcommand{\iter}[2]{#1^{\ing{(#2)}}}

\definecolor{light-gray}{gray}{0.75}

\DeclareMathOperator*{\aud}{\text{a}}
\DeclareMathOperator*{\geo}{\text{g}}

\usepackage[french,serbian,english]{babel}
\usepackage{slashbox}
\graphicspath{{figures/}}

\title{SCATTERING FEATURES FOR MULTIMODAL GAIT RECOGNITION}
\name{Sr\dj{}an Kiti\'c, Gilles Puy, Patrick P\'erez, Philippe Gilberton}
\address{Technicolor, Cesson-S\'evign\'e, France}
\begin{document}
%
\maketitle
\begin{abstract}
We consider the problem of identifying people on the basis of their walk (gait) pattern. Classical approaches to tackle this problem are based on, \emph{e.g.} video recordings or piezoelectric sensors embedded in the floor. In this work, we rely on the acoustic and vibration measurements, obtained from a microphone and a geophone sensor, respectively. The contribution of this work is twofold. First, we propose a feature extraction method based on an (untrained) shallow scattering network, specially tailored for the gait signals. Second, we demonstrate that fusing the two modalities improves identification in the practically relevant open set scenario.   
\end{abstract}
\section{Introduction}
\label{sec:intro}

Identification lies at the heart of many user-defined services, ranging from movie recommendations to online banking. Due to its practical relevance, the problem of identifying people using various biometrics has triggered a significant amount of research in the signal processing and machine learning communities. Traditional means of identification, such as face \cite{jain2011handbook} or speaker \cite{hansen2015speaker,reynolds2008text} recognition, often require active participation in the recognition process, which may be intrusive in many applications. Therefore, a method that can reliably and \emph{passively} identify people is advantageous in such a context.

In this work, we consider human gait as biometrics for identifying people present in a room. A number of approaches to gait-based identification have been proposed in the past, exploiting different signal modalities influenced by walk pattern, \emph{e.g.} based on video \cite{phillips2002gait,hofmann2014tum}, depth \cite{hofmann2014tum} or underfloor accelerometer measurements \cite{bales2016gender}. An appealing modality is structural (\emph{e.g.} floor) vibration induced by walking, and acquired through \emph{geophones} \cite{pan2015indoor}, since it offers several practical advantages over other commonly used types of signals. One of them is increased security - it stems from the fact that there is no simple method (to the authors' knowledge) that can accurately reproduce one's gait in terms of the vibration signal. Another is preservation of privacy as vibration data is usually not considered a confidential, or sensitive information. 
Finally, the proposed setup is simple and cheap
-- typically one geophone is sufficient to monitor a medium-sized room. Unfortunately, geophone measurements are not very rich in content, due to their very limited bandwidth. Currently, geophones are reliably measuring ground vibrations only in the very low frequency range \cite{hons2006transfer}, while the human footstep energy spans up to ultrasonic frequencies \cite{ekimov2007ultrasonic}. Hence, the loss of information is substantial. 

In addition to vibrations (wave propagation in solids), a walking human also produces audible signals, which can be registered by standard microphones and used for identification \cite{geiger2014acoustic,hofmann2014tum}. These have a much wider bandwidth, and, in addition to footsteps, they are also generated due to friction of the upper body (\emph{i.e.} due to leg and arm movements). However, modestly-priced microphones suffer from poor frequency response at very low frequencies, and the measured signals are susceptible to environmental noise, such as speech or music. Therefore, it seems that the vibration and acoustic modality somehow complement each other: while the former is secure, robust and ``senses'' the low-frequency range, the latter carries more information, particularly at high frequencies. The goal of this work is to demonstrate that gait-based recognition using each of the modalities is a viable means of human identification, and that the two can be successfully coupled together in order to boost identification performance.

In the following section, we discuss the physical origin of acoustic and vibration gait measurements. Then, we introduce a feature extraction technique based on the \emph{scattering transform} \cite{anden2014deep} and the specificities of the gait signal. In addition, we propose a simple feature fusion technique to enhance performance when bimodal measurements are available. Finally, we provide open set identification results, obtained from exhaustive experiments on a home-brewed dataset.

\section{Gait signals}
\label{sec:signals}

\begin{figure}
	\centering
	\includegraphics[width=.4\textwidth]{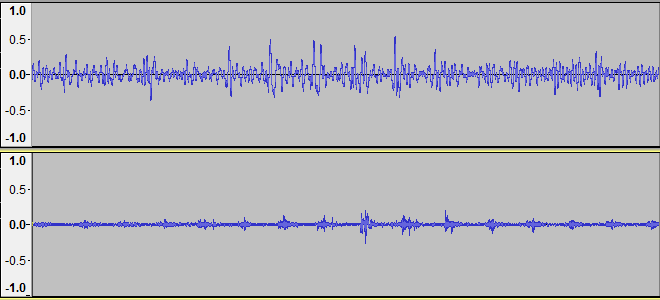}
	\caption{Vibration (top) and acoustic (bottom) recording of a person walking in silence.}\label{figSignals}
\end{figure}

A microphone and a geophone, placed (fixed) at the same location in a room, simultaneously acquire signals of a walking person. Their example outputs are shown in Fig.~\ref{figSignals}: while the two time series are markedly different, the envelope peaks (corresponding to footfalls) are obviously correlated. In fact, the two modalities are linked through latent physical quantity -- (vertical) vibration particle velocity at the impact point -- as described in the remainder of the section. Hereafter, $\vec{r}$ denotes the coordinates of the impact (footfall) point relative to the position of the sensors, $\var{t}$ denotes time and $\omega$ denotes the angular frequency. The hat notation $\hat{\cdot}$ is used to denote the Fourier representation $\mathcal{F}(\cdot)$ of a signal.

Acoustic pressure signal $\hat{x}_{\aud}(\omega,\vec{r}) = \mathcal{F}\left(x_{\aud}(\var{t},\vec{r}) \right)$ can be related to the particle velocity $\hat{v}(\omega)$, as follows \cite{ekimov2006vibration}:
\begin{equation}\label{eqMicTransfer}
	\hat{x}_{\aud}(\omega,\vec{r}) = \hat{h}_{\aud}(\omega,\vec{r}) \hat{v}(\omega) + \hat{e}_{\aud}(\omega) = \hat{g}_{\aud}(\omega,\vec{r})\frac{\hat{v}(\omega)}{\hat{z}(\omega)} + \hat{e}_{\aud}(\omega),
\end{equation}
where $\hat{e}_{\aud}(\omega)$ is the additive noise of the microphone, and $\hat{h}_{\aud}(\omega,\vec{r})$ denotes the microphone transfer function. The latter comprises \emph{specific acoustic impedance} $\hat{z}(\omega)$ (which is a material-related quantity of a medium \cite{fahy2000foundations}) at the impact point, and the (air) impulse response $\hat{g}_{\aud}(\omega,\vec{r})$, relating the impact point and the microphone location. While we may assume that the floor is an isotropic solid -- thus $z(\omega)$ does not change significantly with regards to $\vec{r}$ -- the impulse response $\hat{g}_{\aud}(\omega,\vec{r})$ is influenced to a larger extent by the change in position (this has been empirically verified in \cite{bard2008human}).

Geophone measures the voltage corresponding to the velocity of the proof mass relative to the device case. In the prescribed frequency range, the velocity of the proof mass can be related to the impact point velocity $\hat{v}(\omega)$ \cite{ekimov2006vibration} as
\begin{equation}\label{eqGeoTransfer}
	\hat{x}_{\geo}(\omega,\vec{r}) = \hat{h}_{\geo}(\omega,\vec{r}) \hat{v}(\omega) + \hat{e}_{\geo}(\omega) = S_{\geo} \hat{g}_{\geo}(\omega,\vec{r}) \hat{v}(\omega) + \hat{e}_{\geo}(\omega),
\end{equation}
where $\hat{e}_{\geo}(\omega)$ is the additive noise of the geophone, and $\hat{h}_{\geo}(\omega,\vec{r})$ is the geophone transfer function. Furthermore, $S_{\geo}$ denotes the sensitivity constant, while $\hat{g}_{\geo}(\omega,\vec{r})$ is the impulse response \emph{within the floor} (hence different from $\hat{g}_{\aud}(\omega,\vec{r})$). 

Transfer functions $\hat{h}_{\aud}(\omega,\vec{r})$ and $\hat{h}_{\geo}(\omega,\vec{r})$ (analogously, signals $\hat{x}_{\aud}(\omega,\vec{r})$ and $\hat{x}_{\geo}(\omega,\vec{r})$) are therefore, dependent on $\vec{r}$ - the parameter we cannot control. This is the relative position 
between the walking person and the immobile sensors, 
which thus depends on time $\var{t}$, \emph{i.e.}, $\vec{r} := \vec{r}(\var{t})$. We assume that $\vec{r}(\var{t})$ is a slowly varying function, \emph{i.e.}, 
the impulse responses are (locally) stationary with respect to $\vec{r}$ within a relatively short temporal window, and one can write
\begin{align}
x_{\aud}(\var{t},\vec{r}) \approx x_{\aud}(\var{t}) & = h_{\aud}(\var{t})*v(\var{t}) + e_{\aud}(\var{t}) \; \text{and}\\
x_{\geo}(\var{t},\vec{r}) \approx x_{\geo}(\var{t}) & = h_{\geo}(\var{t})*v(\var{t}) + e_{\geo}(\var{t}), 
\end{align}
where $h_{\aud}(\var{t})$ and $h_{\geo}(\var{t}) $ are time-domain representations of $\hat{h}_{\aud}(\omega) \approx \hat{h}_{\aud}(\omega,\vec{r})$ and $\hat{h}_{\geo}(\omega) \approx \hat{h}_{\geo}(\omega,\vec{r})$, respectively.
	
\section{Feature extraction}
\label{sec:features}

The hypothesis is that the impact velocity $v(\var{t})$ is sufficiently informative to discriminate people. The sensors, however, measure only the bandlimited convolution of $v(\var{t})$ with the corresponding transfer functions. Fortunately, the local stationarity assumption enables us to exploit cancellation property of the so-called \emph{normalized scattering} representation. 

\subsection{Scattering transform}

Scattering tranform is a novel feature extraction method, based on a cascade of wavelet transforms and modulus nonlinearities, bearing some resemblence to convolutional neural networks \cite{bruna2013invariant,mallat2016understanding}. An appealing property of scattering networks is that their filters are pre-defined, hence they require no training. Yet, classifiers using scattering features exhibit almost state-of-the-art performance on several problems, \emph{e.g.} \cite{anden2014deep,bruna2013invariant}. In the following, 
we briefly describe how the scattering transform is computed on the audio signal $x_{\aud}(\var{t})$. 
The features from $x_{\geo}(\var{t})$ are extracted in the same manner.

For a scattering of \emph{order} $\ing{p}$, the features are computed as $S_{\lambda_1 \hdots \lambda_{\ing{p}}}(x_{\aud}(\var{t}),\var{t}) = \phi_T(\var{t}) * U_{\lambda_1 \hdots \lambda_{\ing{p}}} (x_{\aud}(\var{t}),\var{t})$. Here, $\phi_T$ denotes the real-valued lowpass filter of bandwidth $2\pi/T$ (where $T$ is the targeted extent of time-invariance), and $U_{\lambda_1 \hdots \lambda_{\ing{p}}}(\cdot)$ is the so-called the \emph{wavelet propagator}\footnote{For notational convenience, the variables $\var{t}$ and $\omega$ are dropped when the dependence is obvious.}:
\begin{equation}
	U_{\lambda_1 \hdots \lambda_{\ing{p}}} (x_{\aud}) = | \psi_{\lambda_{\ing{p}}} * | \psi_{\lambda_{\ing{p-1}}} * | \hdots  | \psi_{\lambda_1} * x_{\aud} | \hdots |,
\end{equation}
where $\psi_{\lambda_{\ing{i}}} := \psi_{\lambda_{\ing{i}}}(\var{t})$ is a complex analytic wavelet filterbank at $0 < \ing{i} \leq \ing{p}$. The set of scales  $\lambda_{\ing{i}} \in \Lambda_{\ing{i}}$ is chosen such that the filterbank covers the frequency range $[\pi/T, \omega_{\aud}/2]$ ($\omega_{\aud}$ is the sampling frequency), possibly with certain redundancy. The expression above defines the recursion $U_{\lambda_1 \hdots \lambda_{\ing{p}}} (x_{\aud}) = | \psi_{\lambda_{\ing{p}}} * U_{\lambda_1 \hdots \lambda_{\ing{p-1}}} (x_{\aud}) |$, with $U_{\varnothing} (x_{\aud}) = x_{\aud}$ at $\ing{i} = 0$. 

In \cite{anden2014deep}, the authors further refine scattering features by making them nearly invariant to convolution by a filter $h$, when $\hat{h}$ is almost constant on the support of $\psi_{\lambda_{\ing{i}}}$, which we assume to hold in our application. These normalized scattering coefficients are computed as component-wise division
\begin{align}
	& \tilde{S}_{\lambda_1}(x_{\aud}) = \frac{S_{\lambda_1}(x_{\aud})}{\phi_T * |x_{\aud}| + \varepsilon}, \, \varepsilon > 0, \; \text{for} \; \ing{i}=1,\\
	\text{and} \; & \tilde{S}_{\lambda_1 \hdots \lambda_{\ing{i}}} (x_{\aud}) = \frac{S_{\lambda_1 \hdots \lambda_{\ing{i}}} (x_{\aud})}{S_{\lambda_1 \hdots \lambda_{\ing{i}-1}} (x_{\aud})}, \; \text{for} \; \ing{i}>1.
\end{align}
The zero-order coefficients $\tilde{S}_{\varnothing}(x_{\aud}) := S_{\varnothing}(x_{\aud}) = \phi_T * x_{\aud}$ remain unchanged.

Hence, if we independently consider signal segments of duration $\tau$ for which our local stationarity assumption holds, the normalized scattering features should be invariant to filtering by $h_{\aud}$ (accordingly, filtering by $h_{\geo}$ for the geophone signal), and would mostly reflect the behavior of the fingerprint function $v$ in a given bandwidth. 

\subsection{Feature fusion}

When bimodal (microphone and geophone) measurements are available, one can exploit the fact that their effective bandwidths -- frequency ranges for which \emph{SNRs (Signal-to-Noise-Ratio)} is high -- are somewhat complementary. Excluding $\tilde{S}_{\varnothing}(\cdot)$, their respective normalized scattering representations should be complementary as well: the most informative coefficients of each modality appear at scales that do not overlap with one another, except perhaps within a narrow band. Indeed, while the vibration signal $x_{\geo}$ has a very low and narrow frequency range, the audio $x_{\aud}$ is a wideband signal.

\begin{figure}
	\centering
	\includegraphics[width=.23\textwidth]{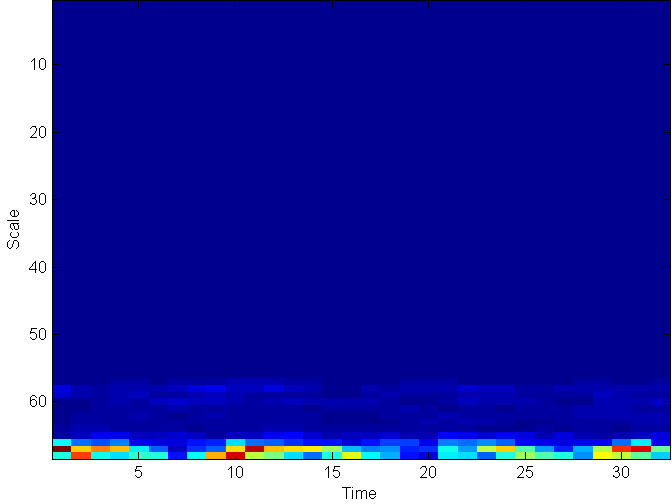}
	\includegraphics[width=.23\textwidth]{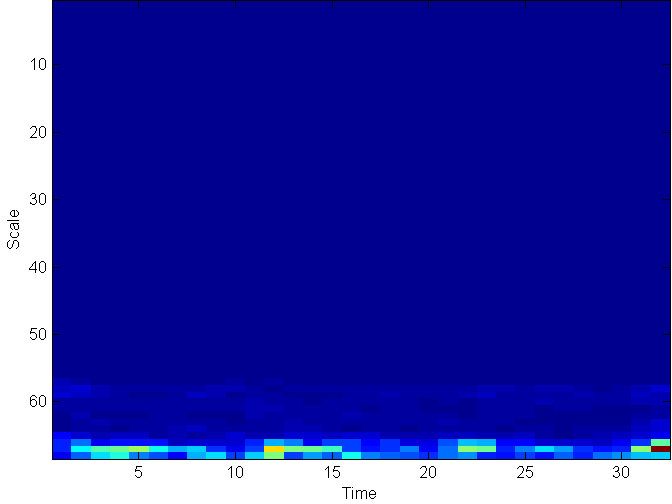}\\
	\includegraphics[width=.23\textwidth]{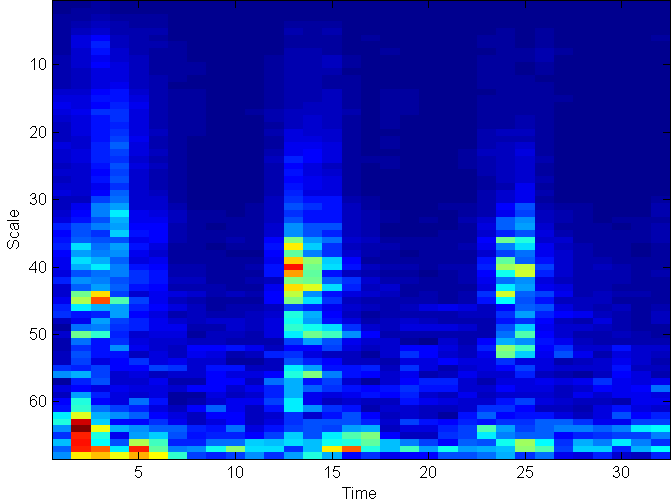}	
	\includegraphics[width=.23\textwidth]{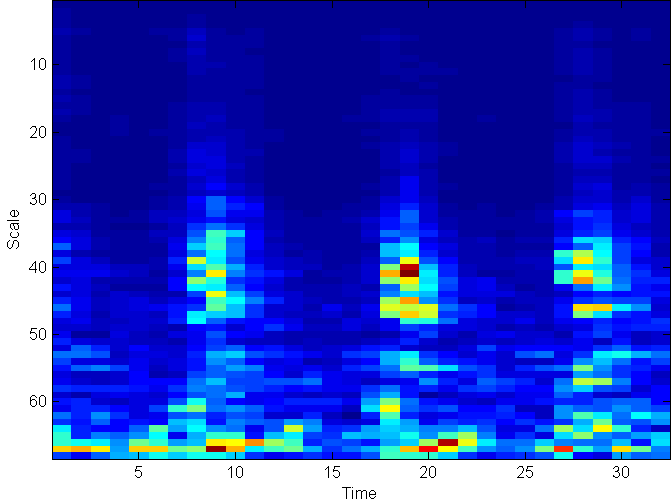}
	\caption{Normalized ($\ing{p}=1$) scattering features for the vibration (top) and audio (bottom) modality, representing the same person, at different time instances (left/right).}\label{figScattering}
\end{figure}

This intuition can be verified in Fig.~\ref{figScattering}, where dark color indicates low magnitude coefficients, and vice-versa. For simplicity, the geophone signal $x_{\geo}$ is upsampled to match the length of the audio signal $x_{\aud}$, thus the feature matrices have the same size. This suggests a simple fusion technique: since the coefficients are nonnegative, one can simply compute a weighted average of the two modalities to obtain a more informative representation, whose (implicit) effective bandwidth is extended. We remark that this is not a pure heuristics, as normalized scattering approach described before places the two representations in the same ``impact velocity feature space''.

The fused scattering $\iter{S_{\cdot}}{f}$ at orders $\ing{i} \geq 1$ is given as
\begin{equation}
	\iter{S_{\lambda_1,\hdots \lambda_{\ing{i}}}}{f} = \alpha(x_{\aud}) \tilde{S}_{\lambda_1,\hdots \lambda_{\ing{i}}}(x_{\aud}) + \alpha(x_{\geo}) \tilde{S}_{\lambda_1,\hdots \lambda_{\ing{i}}}(x_{\geo}),
\end{equation}
with a weight $\alpha(\cdot)$ defined as 
\begin{equation}
\alpha(\cdot) = \left( \max_{ \{ \lambda_{\ing{i}} \in \Lambda_{\ing{i}} \}_{1\leq \ing{i} \leq \ing{p}} } \tilde{S}_{\lambda_1,\hdots \lambda_{\ing{i}}}(\cdot) \right)^{-1},
\end{equation} 
to account for the magnitude disparity among the modalities.

The rows corresponding to zero-order coefficients $\tilde{S}_{\varnothing}(x_{\aud})$ and $\tilde{S}_{\varnothing}(x_{\geo})$ are simply concatenated with the fused ones.

\subsection{Feature postprocessing}

The lowpass filtering by $\phi_T$ makes the output invariant to translations smaller than $T$. It was shown \cite{mallat2016understanding} that the information loss introduced by lowpass filtering is compensated by the higher-order scattering coefficients, with the scattering order $\ing{p}$ predominatelly driven by the signal content \cite{bruna2013invariant,anden2014deep}. 
The rule of thumb is that the larger $T$ is, the higher order the scattering transform should be.
Unfortunately, this significantly increases computational complexity: scattering transform yields a tree-like representation (\emph{cf.} Fig.~2 in \cite{bruna2013invariant}), where each ``path'' $\{\lambda_1,\lambda_2,\hdots \lambda_{\ing{p}} \}_{\lambda_1 \in \Lambda_1,\lambda_2 \in \Lambda_2,\hdots \lambda_{\ing{p}} \in \Lambda_{\ing{p}}}$ needs to be traversed (\emph{i.e.} a full sequence of convolutions needs to be performed) to reach a leaf node. 

As applications enabled by person identification often require real-time processing, our aim is to reduce the computational burden and compute normalized scattering features only up to $\ing{p}=1$ order (``shallow'' scattering network), which implies that $T$ cannot be large. However, features computed from very short signal segments cannot capture temporal dynamics of the gait, which we deem useful for identification. Indeed, the average period of normal walk is about $~1.22$s (two footfalls with the same leg) \cite{ekimov2011rhythm}, and computing sufficiently informative scattering features with $T$ that large is computationally prohibitive. 
While sophisticated classifiers, such as those based on Hidden Markov Models \cite{geiger2014acoustic}, could be used to model the temporal dynamics between successive feature vectors, we opted for a simpler alternative. By inspecting two scattering feature matrices, with the same label but computed at different time instances (Fig.~\ref{figScattering} left and right), one may notice that the main source of variability is due to global temporal offset. This can be easily suppressed by computing the Fourier transform of the scattering matrix across temporal direction, and applying the modulus operator, \emph{i.e.} by discarding the phase. Thus, we extract a segment of duration $\tau > 1.22\text{s} \gg T$, and then postprocess the obtained feature matrix by applying the Fourier modulus row-wise. Since very long segments violate the local stationarity assumption, we set $\tau \approx 1.5$s. Hence, the postprocessing phase introduces additional layer to the first order scattering network.

As suggested in \cite{anden2014deep}, to separate multiplicative signal components and reduce dimensionality, we apply logarithm and \emph{PCA (Principal Component Analysis)} -- or its approximation through \emph{DCT (Discrete Cosine Transform)} -- to postprocessed the feature matrix. The features are standardized (centered and scaled to unit variance) before PCA (DCT).

\section{Results}
\label{sec:results}

While gait recognition attracted considerable amount of research, vibration- and audio-based bimodal identification has not been investigated so far, to the best knowledge of the authors. 
This led us to build our own dataset, by simultaneously recording signals using one ION\textsuperscript{\texttrademark} SM-24 geophone (sampling rate $\approx 1$kHz), and one Samson Meteor\textsuperscript{\textregistered} microphone ($44.1$kHz). The recordings involved $8$ male and $4$ female participants, each recorded during three days, and asked to wear the same type of shoes on (at least) two different days. All recordings were taken in the same room with carpet floor covering. The participants walked the same route $10$ times per day: starting $\sim6$m away, they approached the sensors, and returned to the initial point.

\begin{table}
	\small
	\centering
	\begin{tabular}{|r||c|c|c|c|}
		\hline 
		\backslashbox{$T$}{$\ing{N}$} & 30 & 50 & 100 & 150 \\                 
		
		\hline 
		0.046\text{s} & 23.18\% & 20.00\% & 16.13\%  & 15.08\% \\
		\hline 
		0.093\text{s} & 16.56\% & 15.48\% & 15.96\% & \textbf{13.35\%} \\
		\hline
		0.186\text{s} & 16.67\% & 15.96\% & 15.08\% & 16.13\% \\
		\hline
		0.371\text{s} & 20.00\% & 19.34\% & 19.35\% & 21.64\% \\
		
		\hline 
		\hline          
		
		0.046\text{s} & 26.67\% & 25.81\% & 27.92\%  & 26.80\% \\
		\hline 
		0.093\text{s} & 25.49\% & 23.33\% & 23.47\% & 31.94\% \\
		\hline
		0.186\text{s} & \textbf{20.78\%} & 24.44\% & 29.51\% & 30.03\% \\
		\hline
		0.371\text{s} & 23.87\% & 29.22\% & 28.89\% & 29.03\% \\
		
		\hline 
		\hline            
		
		0.046\text{s} & 19.05\% & 16.67\% & 12.86\%  & 12.09\% \\
		\hline 
		0.093\text{s} & 16.13\% & 12.79\% & 12.38\% & \textbf{10.00\%} \\
		\hline
		0.186\text{s} & 15.96\% & 14.74\% & 13.33\% & 13.33\% \\
		\hline
		0.371\text{s} & 19.15\% & 16.67\% & 18.21\% & 19.68\% \\		
		\hline
	\end{tabular}
	\caption{EER performance of the audio (top), vibration (middle) and fused (bottom) features (lower is better).}
	\label{tabResults}
\end{table}
 
 
Open set identification refers to the case when classes not seen during training may appear in the test phase, and the recognition system needs to label them as ``unknowns''. This type of problem is common in speaker recognition, which shares many traits with gait-based identification (interestingly, in the referenced literature, we found no connection between the two). The gist of current state-of-technology in speaker recognition are variants of \emph{GMM-UBM (Gaussian Mixture Model - Universal Background Model)} framework -- an interested reader may consult \emph{e.g.} \cite{hansen2015speaker,reynolds2008text} -- which we here apply to gait identification. The gait dataset is divided into the ``training'' and ``test'' sets, such that the ``training'' set contains recordings taken on those two days when the participants wore different type of shoes. 
In this way, we ensure that the training data is sufficiently diverse.  The data recorded on the third day constitutes the test set.

We split the training dataset such that the recordings of $6$ randomly chosen individuals are used for training the UBM, and the training data of $3$ among the remaining $6$ (also randomly chosen), is used for the enrollement \emph{cf.} \cite{reynolds2008text}. The test data of these $6$ participants is used in the evaluation phase (thus, there are $3$ unknown persons). This random partitioning is repeated $100$ times, to verify that the results are consistent. 

Normalized scattering, with a redundant Morlet wavelet filterbank, is computed on overlapping signal segments of duration $\tau$ (stepsize $=0.25$s), using the Scatnet toolbox \cite{sifre2013scatnet}. The GMM-UBM system \cite{sadjadi2013msr}, with $64$ Gaussian components, is then fed with the postprocessed scattering features. 

\begin{figure}
	\centering
	\includegraphics[width=0.35\textwidth]{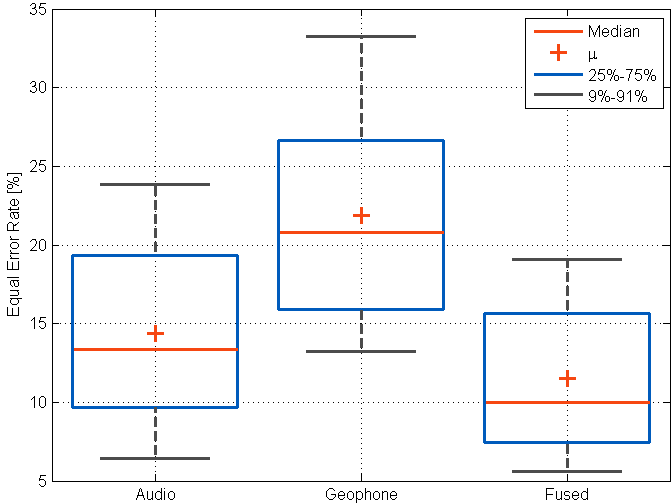}
	\caption{Best performance for each feature type.}\label{figBoxplot}
\end{figure}

Series of experiments is performed by varying the hyperparameters $T$ and $\ing{N}$ (the number of retained DCT coefficients), for each random partition. 
The median results, in terms of \emph{EER (Equal Error Rate)} \cite{hansen2015speaker}, are presented in Table~\ref{tabResults}. Overall, as expected, with the geophone-only features the recognition is somewhat poor. The audio modality performs better, while the fused features perform best, regardless of parameterization. Boxplots for the best-performing parameterizations (boldface values in the table), in Fig.~\ref{figBoxplot}, show that the EERs of the fused representation have the smallest variance. Concerning the choice of time-invariance parameter $T$, the optimal value is between $0.093$s and $0.186$s, which is consistent with average duration of the footfall impact event \cite{ekimov2011rhythm}. The preferred number of features seems to be modality-dependent (\emph{e.g.} richer representations favor larger $\ing{N}$), and may be related to the preset number of GMM components.
 
\section{Conclusion}
\label{sec:conclusion} 
 
We have presented a novel feature extraction approach for person identification based on audio and vibration gait measurements. In a low ambient noise environment, using the audio modality increases recognition accuracy, as demonstrated by the exhaustive experimentation on our bimodal signal dataset. Additionally, we have shown that the two modalities can be fused together to further improve recognition performance. Future work will focus on recognition in adverse conditions, \emph{e.g.} in the presence of auditory noise, and/or several people walking. For the latter, we feel that a body of work on speaker diarization \cite{anguera2012speaker} could be exploited to target such problems. Moreover, bimodal data may offer distinct advantages, both in terms of ``walker diarization'', but also in terms of robustness to ambient noise, since the two modalities are usually not simultaneously affected by the same noise source. Finally, in this work we opted for (deterministic) scattering feature extraction, due to the size of our training dataset. If this is not a limiting factor, recent trends in machine learning suggest that a deep neural network may achieve superior performance.

\bibliographystyle{IEEEbib}
\bibliography{main}

\end{document}